\documentclass[usenatbib]{mn2e}
\pdfoutput=1
\usepackage{graphicx}
\usepackage{mathptmx}
\usepackage{pdfsync}
\usepackage{amsmath}
\usepackage{amssymb}
\usepackage{aas_macros}
\usepackage[usenames,dvipsnames]{color}
\usepackage[a4paper,total={17.8cm,24.0cm},centering]{geometry}
\usepackage[pdfpagelabels]{hyperref}
\hypersetup{pdfauthor={Michael J. Williams et al.},pdftitle={Secular
        evolution in action: Central values and radial trends in the
stellar populations of boxy bulges},breaklinks=true}

\newcommand{\refsec}[1]{Section~\ref{#1}}
\newcommand{\reffig}[1]{Fig.~\ref{#1}}

\newcommand{\kms}{km\,s$^{-1}$}

\newcommand{\oiii}{[O\,\textsc{iii}]}
\newcommand{\nni}{[N\,\textsc{i}]}
\newcommand{\ZH}{[\ensuremath{\mathrm{Z}/\mathrm{H}}]}
\newcommand{\aFe}{[\ensuremath{\alpha/\mathrm{Fe}}]}
\newcommand{\mgb}{\ensuremath{\mathrm{Mg}\,b}}

\newcommand{\halpha}{H$\alpha$}
\newcommand{\hbeta}{\ensuremath{\mathrm{H}\beta}}

\title[Stellar populations of boxy bulges]{Secular evolution in action:
central values and radial trends in the stellar populations of boxy
bulges}

\author[Michael J. Williams et al.]{Michael J.
    Williams$^{1}$\thanks{Email: williams@mpe.mpg.de}, Martin Bureau$^2$
    and Harald Kuntschner$^3$
\\$^1$Max Planck Institute for Extraterrestrial Physics, PO Box 1312,
    Giessenbachstr., 85741 Garching bei M\"unchen, Germany
\\$^2$Sub-Department of Astrophysics, University of Oxford, Denys
Wilkinson Building, Keble Road, Oxford OX1 3RH, UK
\\$^3$European Southern Observatory, Karl-Schwarzschild-Str. 2, D-85748
Garching bei M\"unchen, Germany
}

\begin{document}

\date{Accepted 2012 September 13. Received 2012 September 08; in original form 2012 July 31}

\pagerange{\pageref{firstpage}--\pageref{lastpage}} \pubyear{2012}

\maketitle
\label{firstpage}

\begin{abstract} 
We determine central values and radial trends in the stellar populations
of the bulges of a sample of 28 edge-on S0--Sb disk galaxies, 22 of
which are boxy/peanut-shaped (and therefore barred). Our principal
findings are the following. (1) At a given velocity dispersion, the
central stellar populations of galaxies with boxy/peanut-shaped bulges
are indistinguishable from those of early-type (elliptical and S0)
galaxies. Either secular evolution affects stellar populations no
differently to monolithic collapse or mergers, or secular evolution is
not important in the central regions of these galaxies, despite the fact
that they are barred. (2) The radial metallicity gradients of
boxy/peanut-shaped bulges are uncorrelated with velocity dispersion and
are, on average, shallower than those of unbarred early-type galaxies.
This is qualitatively consistent with chemodynamical models of bar
formation, in which radial inflow and outflow smears out pre-existing
gradients.
\end{abstract}


\section{Introduction}
\label{sec:dbs:intro}

Among the observational clues and tests for theories of galaxy formation
and evolution are the well-known empirical correlations between the
optical absorption line strengths of early-type galaxies (ellipticals
and S0s, ETGs hereafter) and their dynamical or photometric properties
\citep[e.g.][]{Terlevich:1981,Bender:1993,Kuntschner:2000}. For
increasing stellar velocity dispersion or mass, central and global
observations of the line strengths (and therefore stellar populations)
of ETGs show them to be older and more metal-rich, and more abundant in
$\alpha$-elements relative to Fe
\cite[e.g.][]{Thomas:2005,Kuntschner:2010}. Radial trends in stellar
populations are a powerful additional discriminant. Monolithic collapse
models (e.g.\ \citealt*{Eggen:1962};
\citealt{Larson:1974,Carlberg:1984}) predict steep negative metallicity
gradients of $-0.35 < \Delta\ZH < -1.0$.\footnote{We follow convention
    by characterizing radial gradients with linear fits in $\log R$, and
    using the notation $\Delta\mathrm{X} \equiv
\partial\,\mathrm{X}/\partial\log R$, where X is \ZH, \aFe\ or
$\log(\mathrm{age/Gyr})$.} Hierarchical models predict that these
gradients become generally shallower as they are diluted by mergers
\citep[e.g.][]{White:1980,Di-Matteo:2009}.

Observations of stellar population gradients in ETGs
\citep[e.g.][]{Davies:1993,Mehlert:2003,Kuntschner:2006,Spolaor:2009,Kuntschner:2010,Roediger:2011}
show that, on average, $\Delta\log(\mathrm{age/Gyr}) \approx 0$ and
$\Delta\aFe \approx 0$. However, \mbox{$\Delta\ZH \approx -0.2\pm0.1$},
and below a central velocity dispersion of $\approx 150$\,\kms\
(equivalent to a dynamical mass $\approx 3 \times
10^{10}\,\mathrm{M}_\odot$), there is some evidence of a correlation:
more massive systems have steeper negative metallicity gradients
\citep[e.g.][]{Spolaor:2009,Kuntschner:2010}. Above this characteristic
mass, the correlation between gradient and velocity dispersion or mass
disappears, but the average metallicity gradient is still negative. 

Analysis of the stellar populations of the bulges of spiral galaxies is
more challenging than in ETGs: bulges are fainter and embedded in disks,
and they may have complicated or ongoing star formation histories,
nebular emission, and significant dust absorption. Nevertheless, the
consensus is that the stellar populations of bulges in S0-Sbc disk
galaxies are very similar to those in ellipticals and S0s. At a given
velocity dispersion, their line strengths and the implied stellar
populations are, on average, the same as those of earlier types
\citep{Proctor:2002,Thomas:2006,Falcon-Barroso:2006,MacArthur:2009}.
In combination with dynamical and structural similarities, this is
evidence that the bulges of these galaxies formed in a similar way to
ETGs. However, the good agreement between bulges and ETGs breaks down in
the bulges of late-type spirals. For example, \cite{Ganda:2007} observed
that the bulges of a sample of 18 Sb--Sd galaxies have smaller \mgb\
indices and larger \hbeta\ indices at a given stellar velocity
dispersion $\sigma$. The differences may be evidence that these bulges
formed or are currently affected by different processes to ETGs and the
bulges of earlier-type disks, the most likely being the secular and
internal rearrangement of disk material \citep{Kormendy:2004}.

The radial behaviour of the stellar populations of bulges in spiral
galaxies is relatively uncertain. The major studies are \citet[long-slit
observations of the bulges of 38 galaxies ranging from S0 to
Sbc]{Moorthy:2006}, \cite{Falcon-Barroso:2006} and \citet[SAURON
integral field observations of the bulges of 24 Sa
galaxies]{Peletier:2007}, \citet[SAURON integral-field observations of
the bulges of 18 Sb--Sd galaxies]{Ganda:2007}, \citet[minor-axis
long-slit observations of the bulges of 30 edge-on S0 to Sc
galaxies]{Jablonka:2007}, \citet[long-slit observations of the bulges 18
S0--Sbc galaxies]{Morelli:2008} and \citet[long-slit observations of the
bulges of 8 Sa--Sd galaxies]{MacArthur:2009}. Among the areas of
agreement of these studies is the finding that most bulges have negative
metallicity gradients. Compared to ETGs, however, there is rather more
scatter in these gradients and more diversity in the structure of the
radial profiles. This observation brings us on to the subject of bars
(and, by implication, boxy/peanut-shaped bulges), since they may be the
origin of that diversity. 

\defcitealias{Perez:2011}{PSB11}

Simulations of the chemodynamical evolution of barred disk galaxies find
that the bars drive an inflow of gas within corotation and outflow
beyond \citep[e.g.][]{Friedli:1994,Friedli:1998}. This radial transport
naturally flattens pre-existing population gradients. Moreover,
simulations by \cite{Wozniak:2007} predict local minima of stellar age
at the ends of bars. Observational tests of these predictions are
crucial, because bars are found in two-thirds of disk galaxies and
likely play a significant role in transforming the stellar populations
of their hosts. Perez et al.\ have made the first systematic attempt to
study radial stellar population trends in barred galaxies \citep[][PSB11
hereafter]{Perez:2007,Perez:2009,Perez:2011}. They used optical spectra
taken with long slits oriented along the bar major axes of 20 face-on
and moderately-inclined barred S0--Sb galaxies.

This Letter studies central values and radial trends in the stellar
populations of bulges of barred galaxies by determining absorption line
strengths and single stellar population (SSP) equivalent ages, \ZH\ and
\aFe. We use the \cite{Bureau:1999} sample of edge-on disk galaxies. The
majority (22/28) of the galaxies in this sample host bulges that are
boxy or peanut-shaped, i.e.\ bars viewed in projection
\citep[e.g.][]{Combes:1990,Kuijken:1995,Bureau:1999,Chung:2004,Kormendy:2004}.
In \refsec{sec:dbs:data} we describe the sample, observations and data
reduction. In \refsec{sec:dbs:central} we discuss the central and global
stellar populations, and in \refsec{sec:dbs:radial} we discuss the
radial trends and gradients. Full results (i.e.\ complete radial
profiles) for the stellar kinematics, gas kinematics, Lick indices, and
SSP-equivalent population properties are presented in
\cite{Williams:2011a}. 

\section{Observations and data reduction}
\label{sec:dbs:data}

Our sample galaxies are the 28 edge-on galaxies presented in
\cite{Bureau:1999} and \cite{Chung:2004}. Those works describe the
sample and observations used here in more detail, but in brief half of
the galaxies are S0s, half are spirals, and 22/28 host a boxy or
peanut-shaped bulge. We took long-slit spectra using the Double Beam
Spectrograph (DBS) on the 2.3\,m telescope at Siding Springs
Observatory. In the instrumental setup used, the DBS takes two spectra,
each covering a range of $\approx 1000$\,\AA\@. The blue arm was centred
on the continuum absorption features around the \mgb\ triplet. The red
arm was centred on the \halpha\ emission line when detected or moved to
the Calcium triplet. The red spectra therefore suffer from emission or
strong atmospheric absorption and cover a wavelength range where stellar
population models are poorly constrained. The blue arm spectra are
therefore the sole focus of this work.

We reduced the data from the blue arm using standard IRAF long-slit
techniques, yielding a flat-fielded, wavelength-calibrated,
sky-subtracted long-slit spectrum along the major axis of each galaxy.
Cumulative exposure times ranged from 200 to 400 minutes per object. The
reduced two-dimensional spectra were linearly binned in wavelength and
cover the spectral range 4755--5710\,\AA\@. The spectral resolution of the
instrument is 1.2\,\AA\ full-width half-maximum (equivalent to
$\sigma_\mathrm{inst} \approx 30$\,\kms{} at 5150\,\AA{}). The spatial
axis has a pixel scale of 0.91\,arcsec pixel$^{-1}$ and covers a maximum
of 5\arcmin.

\cite{Chung:2004} have already presented absorption line stellar
kinematics for these data, but we rederived them for the present work
for two reasons. Firstly, we need to spatially bin to a higher
signal-to-noise ratio ($S/N$) to better measure absorption line
strengths. Secondly, more powerful and flexible analysis libraries and
codes are now available, which allow us to treat emission lines more
precisely \citep{Cappellari:2004,Sarzi:2006}. Outside the central few
pixels, where the $S/N$ was already large, we spatially binned each
spectra to a $S/N \ge 30$~per~\AA\@.

We extracted stellar and gas kinematics using identical techniques to
those described in \cite{Williams:2011}, i.e.\ using penalized pixel
fitting (PPXF; \citealt{Cappellari:2004}) and GANDALF
\citep{Sarzi:2006}. As in \cite{Williams:2011}, we used a subset of 88
stars from the MILES library of 985 observed stellar spectra
\citep{Sanchez-Blazquez:2006} as templates. We then used GANDALF to
clean any \hbeta, \oiii\ and \nni\ emission, which are the three main
nebular emission features present in our wavelength range. The \hbeta\
emission line occurs in the central passband of the \hbeta\ absorption
Lick index, while \oiii\ and \nni\ occur in the continuum passbands of
the Fe5015 and \mgb\ indices, respectively. We imposed the kinematics of
the \oiii\ emission line on the \hbeta\ and \nni\ lines. This
simplification is justified for our limited goals (cleaning emission)
because (1) in the high $S/N$ galaxies in our sample, where the
kinematics of the \hbeta\ emission line can be reliably constrained,
there is no evidence that they differ systematically from those of the
\oiii\ line, and (2) there is no evidence that the \hbeta\ and \oiii\
emission line kinematics differ by an amount greater than their
uncertainties in similar analyses of either the SAURON ETG sample
\citep{Sarzi:2006} or the SAURON early-type (Sa/Sb) spiral bulge sample
\citep{Falcon-Barroso:2006}.

Using the stellar kinematics and emission-cleaned spectra, we use the
method of \cite{Kuntschner:2006} to measure the strengths of the
absorption lines present in our data (H$\beta$, Fe5015, Mg\,$b$, Fe5270,
Fe5335 and Fe5406) in the Lick/IDS system
\citep{Burstein:1984,Worthey:1994,Trager:1998}. We compare these Lick
index measurements to an interpolated grid of \cite*{Thomas:2003} SSP
models, yielding SSP-equivalent luminosity-weighted ages, metallicities
\ZH\ and $\alpha$-element enhancement \aFe\ \citep[using the method
of][]{Proctor:2004}. We exclude the Fe5270 or Fe5335 feature (depending
on the galaxy redshift) because of bad columns on the CCD of the blue
arm of the DBS\@. We exclude Fe5406 in the most distant galaxy
(ESO~597-G036) because that line is redshifted into the strong
[O\,\textsc{i}] sky line at 5577\,\AA\@. 

\section{Central line strengths and SSP properties}
\label{sec:dbs:central}

\begin{figure} 
    \includegraphics[width=8.6cm]{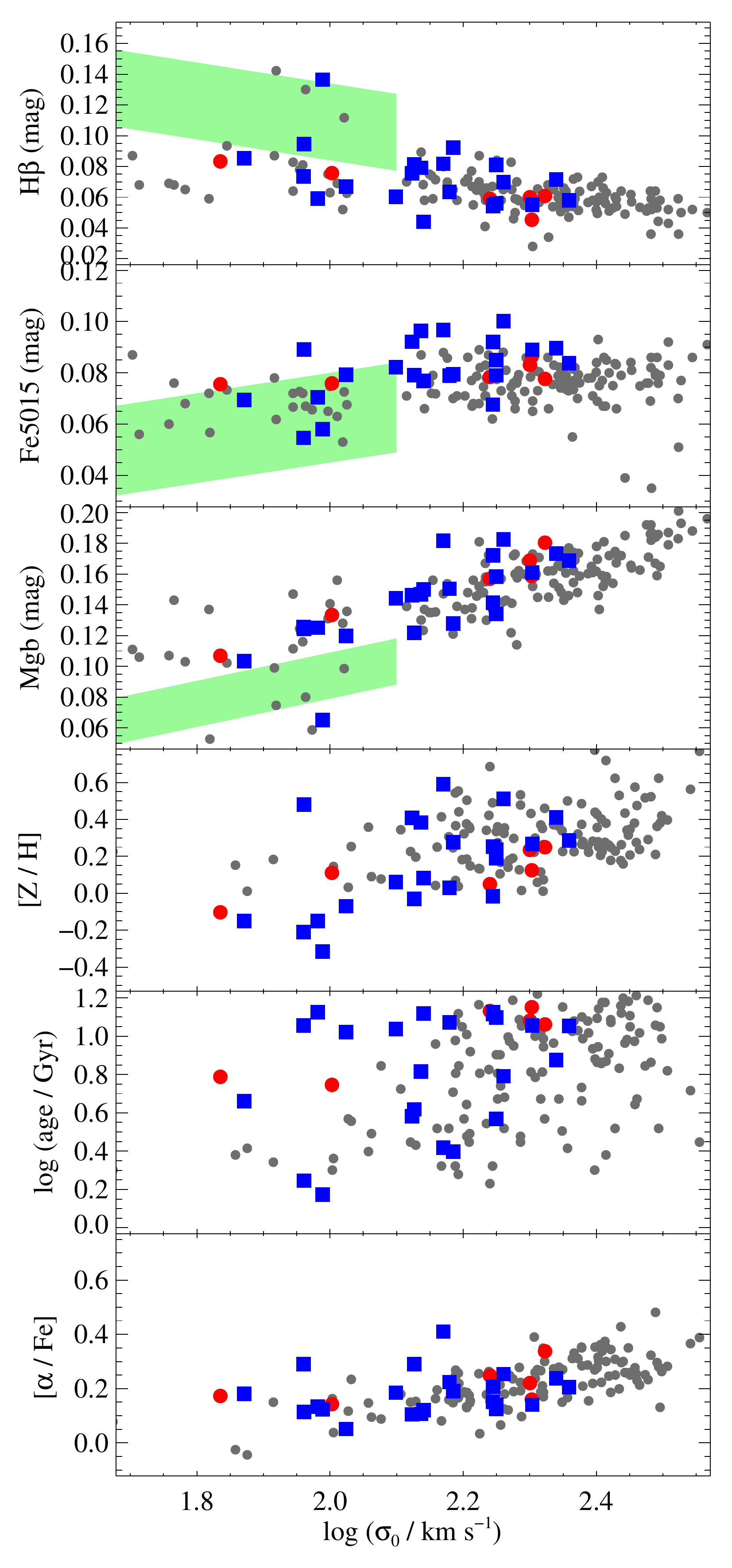}
\caption{Central \hbeta, Fe5015 and \mgb\ Lick indices and luminosity-weighted
    SSP-equivalent population parameters as a function of central
    stellar velocity dispersion. Indices are measured in magnitudes,
    i.e. $-2.5 \log_{10} (1 - I/\Delta I)$, where $I$ is index value and
    $\Delta I$ is the width of its bandpass, both of which are measured
    in angstrom. Large colored symbols are from our sample; blue squares
    are our boxy bulges, and red circles our round bulges. Error bars
    are omitted for clarity; the typical uncertainties are $\pm0.1$\,mag
    in \hbeta, Fe5015 and \mgb\, and $\pm0.1$\,dex in \ZH, log(age/Gyr)
    and \aFe. The smaller gray points are comparison data taken from the
    literature. \emph{Top two panels}: Gray circles are ETGs from
    \protect\cite{Sanchez-Blazquez:2006a} and SAURON
    \protect\citep{Kuntschner:2006}. The shaded green regions are the
    regions populated by the bulges of late-type (Sb--Sd) spirals in the
    sample of \protect\cite{Ganda:2007}. \emph{Bottom three panels}: The
    gray circles are ETGs from \protect\cite{Thomas:2005}, which,
    like ours, use the \protect\cite{Thomas:2003} SSP
models.\label{fig:central}}
\end{figure}

The smallest aperture from which it is meaningful to extract data is set
by the seeing limit of the observations and the width of the slit, i.e.
$3\arcsec \times 1.8\arcsec$. We refer to the average quantities within
this aperture as `central'.

In \reffig{fig:central} we show, as a function of central stellar
velocity dispersion $\sigma_0$, the central values of three
representative Lick indices and the SSP-equivalent age, $\ZH$, and
$\aFe$ of our sample galaxies. We also show comparison data for ETGs. We
find no evidence that the central populations of our barred disk
galaxies differ from those of ETGs at a given velocity dispersion.

Almost all of these galaxies are barred, and therefore good candidates
for observing the effects of secular evolution in their stellar
populations. At least in terms of their central values, however, the
stellar populations of these barred S0--Sb disk galaxies are no
different to those of ETGs (or indeed the six unbarred disk
galaxies in our own sample). This implies that either secular evolution
does not affect the stellar populations of the centers of bulges of
barred S0--Sb galaxies, or its effects are no different to those of
monolithic collapse and mergers, the putative formation mechanisms of
ETGs. We note that this result is consistent with the findings of
\cite{Thomas:2006}, who studied the central stellar populations of
spiral bulges across a broad range of Hubble types, but did not consider
the role of bars in particular. Since our sample is restricted to S0--Sb
galaxies, our result does not necessarily contradict the apparent
observation of secular evolution effects on the central stellar
populations of later-type (Sb--Sd) bulges by \cite{Ganda:2007}.

\section{Radial metallicity gradients}
\label{sec:dbs:radial}

\begin{figure} \includegraphics[width=8.6cm]{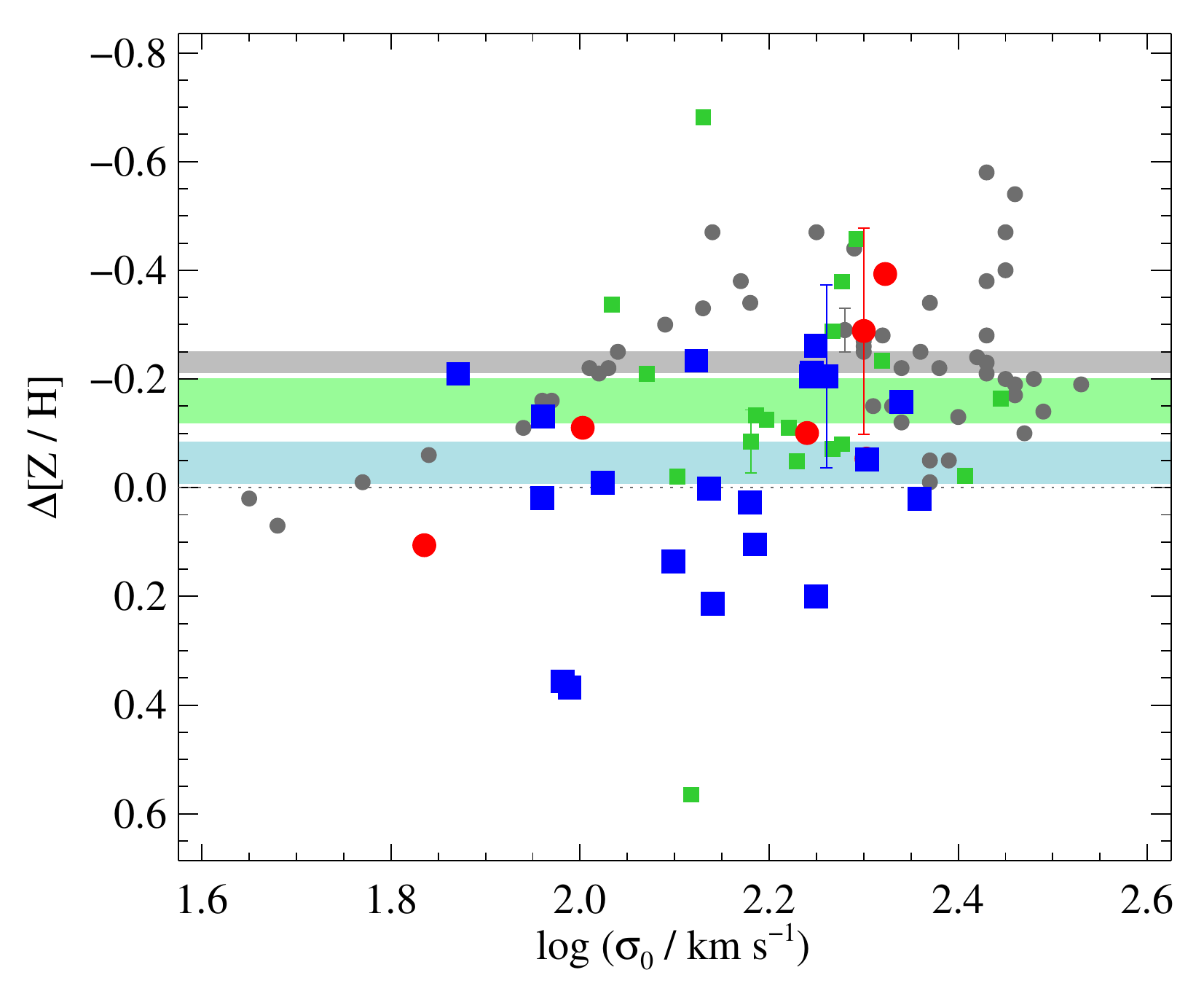} \caption{Radial
        \ZH\ gradients as a function of central velocity dispersion.
        Large symbols are from our sample; blue squares are our boxy
        bulges, and red circles are out round bulges. For clarity, the
        median uncertainties are shown as error bars only on a
        representative data point. Smaller symbols are comparison data;
        green squares are barred S0--Sb galaxies from
        \citetalias{Perez:2011}, gray circles are ETGs from
        \protect\cite{Spolaor:2010a}. The thick lines are the mean
        $\Delta\ZH$ for the samples: pale blue for our boxy bulges, pale
        green for the \citetalias{Perez:2011} barred galaxies, and gray
        for the \protect\cite{Spolaor:2010a} ETGs. The thickness of
    these lines is is the uncertainty on the mean. \label{fig:sigmad}}
\end{figure}

To measure $\Delta\ZH$, we fit a straight line to \ZH\ as a function of
$\log R$. We use the full radial extent of the data, which cover, on
average, the inner $\approx$30\,arcsec, i.e.\ the bulge of these local
galaxies. We exclude all points \emph{inside} the seeing limit, which we
considered separately in the previous section. We plot $\Delta\ZH$ as a
function of central velocity dispersion in \reffig{fig:sigmad}. The data
for 2 of our 28 galaxies were only of sufficient $S/N$ to measure the
central populations, so are excluded from our gradient analysis. For
comparison, we show the \ZH\ and \aFe\ gradients from a catalogue of
ETGs assembled by \cite{Spolaor:2010a}. The original sources of these
data are \cite{Proctor:2003}, \cite{Brough:2007}, \cite{Reda:2007},
\cite{Sanchez-Blazquez:2007}, \cite{Spolaor:2008}, and
\cite{Spolaor:2010a}. All are based on long-slit observations, and all
sources except \cite{Proctor:2003} use the \cite{Thomas:2003} SSP
models. We note in any case that, while the choice of SSP models may
affect the absolute stellar population parameters derived, the radial
gradients are unlikely to be strongly affected \citep{Kuntschner:2010}.

From \reffig{fig:sigmad} we conclude the following. (1) There is no
evidence that $\Delta\ZH$ in our boxy bulges is correlated with velocity
dispersion. (2) The boxy bulges of our sample of barred galaxies have
shallower metallicity gradients than those of ETGs, both on average and
at a given velocity dispersion. The mean value of $\Delta\ZH$ for the
boxy and peanut-shaped bulges is $-0.06\pm0.04$ and there are several
cases of positive metallicity gradients. In contrast, the mean
$\Delta\ZH$ of the \cite{Spolaor:2010a} catalogue of ETGs is
$-0.23\pm0.02$. These results are qualitatively consistent with the
simulations of \cite{Friedli:1994}, who find that outflows and inflows
in barred galaxies make pre-existing radial gradients less steep. Since
the pre-existing age and \aFe\ gradients of ETGs and bulges are, on
average, flat, it is not a surprise that, on average,
$\Delta(\mathrm{age/Gyr}) = 0$ and $\Delta\aFe = 0$ for our sample as
well. We omit these results from \reffig{fig:sigmad}, but the results
for age and \aFe\ profiles are presented and fully discussed in
\cite{Williams:2011a}.

One may reasonably worry that line-of-sight effects in our sample of
edge-on barred galaxies are responsible for some or all of the
flattening of their radial gradients. While we cannot quantify this
effect, we argue that it must be small for two reasons. (1) There is no
systematic difference between the radial gradients of our 14 S0s
(edge-on galaxies largely free of dust) and 14 spirals (edge-on galaxies
with prominent dust lanes). This suggests the role of dust is small. (2)
As shown in \reffig{fig:sigmad}, we see a similar result --- shallower
gradients in barred galaxies --- in data taken from the bulges of a
face-on sample of barred galaxies that cannot suffer from line-of-sight
flattening \citepalias{Perez:2011}. The statistical significance of the
difference between metallicity gradients in the \citetalias{Perez:2011}
galaxies and unbarred ETGs is admittedly weak (the mean $\Delta\ZH$ for
the \citetalias{Perez:2011} sample is $-0.15\pm0.04$), but the strong
and clear correlation between $\Delta\ZH$ and $\sigma$ seen in unbarred
ETGs with $\log(\sigma/\mathrm{km\,s^{-1}}) < 2.2$ \citep{Spolaor:2010a}
is totally absent from both our sample of boxy bulges and the
\citetalias{Perez:2011} barred galaxies. 

\section{Discussion}

In summary, we have shown that the stellar populations at the very
centres of the bulges found in barred S0--Sb galaxies do not differ from
those of ETGs of the same velocity dispersion. On larger physical
scales, however, these bulges do differ from ETGs: they lack the
correlation between metallicity gradient and velocity dispersion found
in ETGs, and their average stellar population metallicity gradients are
shallower than unbarred ETGs with the same velocity dispersion.

It is clear that the role of bars in transforming the stellar
populations of disk galaxies is a significant gap in our understanding
of galaxy evolution. This work suggests several possible avenues for
further study. Simulations should attempt to go beyond the qualitative
statement that bars should make pre-existing abundance gradients
shallower, either by drawing their initial abundance profiles from
samples such as \cite{Spolaor:2010a} and \cite{Kuntschner:2010}, in
which $\Delta\ZH$ is correlated with velocity dispersion, or by using
full cosmological simulations with sufficient resolution to capture
bar-driven evolution. Simulations should also make predictions of the
vertical gradients of barred galaxies as a function of radius
\citep[e.g.][]{Friedli:1998,Williams:2011}. Integral-field observations
of the boxy/peanut-shaped bulges of edge-on galaxies are the ideal
observational tests of such predictions. Observations of face-on
galaxies are complementary to this work, and integral-field data raise
the possibility of looking for anticipated azimuthal variations in
stellar populations, such as local minima in age at the bar ends
\citep{Wozniak:2007,Perez:2007}.

\section*{Acknowledgements}

We thank Isabel P\'erez and Patricia Sanchez-Bl\'azquez for generously
sharing comparison data used in the construction of \reffig{fig:sigmad},
and Alfonso Aragon-Salamanca and Roger Davies for useful discussions.
MJW acknowledges with gratitude the hospitality of Malcolm Bremer and
the University of Bristol Astrophysics Group. MB acknowledges support
from the STFC rolling grant `Astrophysics at Oxford' PP/E00114/1 and
ST/H002456/1.

\label{lastpage}

\end{document}